\documentclass{article}

\usepackage[T2A]{fontenc}
\usepackage[utf8]{inputenc}
\usepackage[english]{babel}
\usepackage{graphicx}

\author{\bfseries Andrey Vasilyev}
\title{\bfseries Integral Property of Laplace Operator}  
\date{Retired from State Optical Institute, Saint Petersburg, Russia
e-mail <andrey@wavemech.org>}

\begin{document}

\maketitle

\begin{abstract}
\label{abstract}

Relations have been derived which establish connection between a scalar or a vector functions and the integral of Laplace operator of these functions (the integral property of Laplace operator). The integral property of Laplace operator was employed to obtain relations expressing the electric field in terms of the charge gradient. It is demonstrated that these relations define the same field which is described by the well known classical expressions for the electric field. It is shown that these proposed expressions represent actually a particular case of the general theory of generalized functions. This approach is illustrated by calculation of an electric field performed with the proposed formalism.
 
\bigskip

{\bfseries Key words:} Laplace operator, Green's theorem, electric field, scalar potential, delta-function.

\end{abstract}

PACS: 41.20.Cv

MSC: 78A25, 78A30


\section{Integral property of Laplace operator}

Consider a scalar, $\Phi(\mathbf r)$, or a vector, $\mathbf V(\mathbf r)$, functions, single-valued and continuous, which have continuous second-order partial derivatives and fall off at infinity faster than $1/r$. We first find the expressions relating these functions to Laplace operator of these functions. To do this,we invoke the Green's theorem (cf., e.g., \cite{1}, 
item 5,6-1)

\begin{equation}
\label{1.1}
\int\limits_V(\Psi\nabla^2\Phi-\Phi\nabla^2\Psi)dV=
\oint\limits_Sd\mathbf S(\Psi\nabla\Phi-\Phi\nabla\Psi).
\end{equation} 

Next we use the approach outlined in monograph \cite{2}, Ch. 1. For function $\Psi$ we use the expression $1/|\mathbf r-\mathbf r'|$. The integration will be performed over the $\mathbf r'$ coordinate space.

Because the integrand $\Phi$ in surface integrals falls off faster than $1/r$, and integration is run over the whole space, surface integrals vanish. In this case Eq. (\ref{1.1}) acquires the form

\begin{equation}
\label{1.2}
\int\frac{1}{|\mathbf r-\mathbf r'|}\nabla^2_{\mathbf r'}\Phi(\mathbf r')dV'=
\int\Phi(\mathbf r')\nabla^2_{\mathbf r'}\frac{1}{|\mathbf r-\mathbf r'|}dV',
\end{equation}
in which integration of $dV'$ is performed over the $\mathbf r'$ coordinates.

As is well known, the Laplacian transforms as 
$\nabla^2_{\mathbf r}\frac{1}{|\mathbf r-\mathbf r'|}=
-4\pi\delta(\mathbf r-\mathbf r')$ 
 (cf., \cite{2}, Ch.~1; \cite{3}, Ch.~1). One feature should be stressed here, because we shall make use of it in the future. The gradients 
$\nabla_{\mathbf r}$ and $\nabla_{\mathbf r'}$ of the function 
$\frac{1}{|\mathbf r-\mathbf r'|}$ differ in sign

\begin{equation}
\label{1.3}
\nabla_{\mathbf r}\frac{1}{|\mathbf r-\mathbf r'|}=
-\frac{\mathbf r-\mathbf r'}{|\mathbf r-\mathbf r'|^3}, \qquad
\nabla_{\mathbf r'}\frac{1}{|\mathbf r-\mathbf r'|}=
\frac{\mathbf r-\mathbf r'}{|\mathbf r-\mathbf r'|^3}.
\end{equation}
Nevertheless, the Laplacians 
 $\nabla^2_{\mathbf r}\frac{1}{|\mathbf r-\mathbf r'|}$ and 
$\nabla^2_{\mathbf r'}\frac{1}{|\mathbf r-\mathbf r'|}$ are equal to the same quantity, $-4\pi\delta(\mathbf r-\mathbf r')$.

Indeed, the expressions 
$\nabla^2_{\mathbf r}\frac{1}{|\mathbf r-\mathbf r'|}$ and 
$\nabla^2_{\mathbf r'}\frac{1}{|\mathbf r-\mathbf r'|}$ are zero everywhere except at the point $\mathbf r=\mathbf r'$. To determine the magnitude of the first expression at point $\mathbf r=\mathbf r'$, we follow the well-known procedure of integrating this expression over the volume of a small sphere  
$R_0\to 0$ centered at the point $\mathbf r=\mathbf r'$. Here the coordinate 
$\mathbf r'$ will remain constant. Denoting 
$|\mathbf r-\mathbf r'|=R$ and $\mathbf r-\mathbf r'=\mathbf R$ we apply Gauss’s theorem to come finally to

$$
\int\limits_V\nabla_{\mathbf r}^2\frac{1}{R}dV=
\oint\limits_S\nabla_{\mathbf r}\frac{1}{R}\cdot d\mathbf S=
-\int\frac{1}{R_0^2}\mathbf n_{R}R_0^2d\Omega\mathbf n_{S}=-4\pi.
$$
Here $\mathbf n_{R}$ is a unit vector in the direction of vector $\mathbf R$,  
$\nabla_{\mathbf r}\frac{1}{R}=-\frac{\mathbf R}{R^3}=-\frac{1}{R^2}\mathbf n_{R}$,
 (see (\ref{1.3})). Vector $\mathbf n_{S}$ is a unit vector of the $S$ surface normal; $\mathbf n_{R}$ and $\mathbf n_{S}$ coincide in direction.

Calculate now the expression 
$\nabla^2_{\mathbf r'}\frac{1}{|\mathbf r-\mathbf r'|}$ at point 
$\mathbf r=\mathbf r'$. We follow now the procedure outlined above, the only difference being that the coordinate $\mathbf r$ is now assumed to be constant. This leads us to 

$$
\int\limits_V\nabla_{\mathbf r'}^2\frac{1}{R}dV=
\oint\limits_S\nabla_{\mathbf r'}\frac{1}{R}\cdot d\mathbf S=
\int\frac{1}{R_0^2}\mathbf n_{R}R_0^2d\Omega\mathbf n_{S},
$$
because, as follows from (\ref{1.3}) 
$\nabla_{\mathbf r'}\frac{1}{R}=\frac{\mathbf R}{R^3}=\frac{1}{R^2}\mathbf n_{R}$. In this case, however, vectors $\mathbf n_{R}$ and $\mathbf n_{S}$, rather than coinciding in direction, are directed oppositely (indeed, vector $\mathbf n_{R}$ is pointed now to point $\mathbf r$, while vector $\mathbf n_{S}$ is directed away from point $\mathbf r$). Therefore this expression again becomes equal to $-4\pi$. This yields 

\begin{equation}
\label{1.4}
\nabla^2_{\mathbf r}\frac{1}{|\mathbf r-\mathbf r'|}=
\nabla^2_{\mathbf r'}\frac{1}{|\mathbf r-\mathbf r'|}=
-4\pi\delta(\mathbf r-\mathbf r').
\end{equation}
As a result, expression (\ref{1.2}) assumes the form we expected:

\begin{equation}
\label{1.5}
\int\frac{1}{|\mathbf r-\mathbf r'|}\nabla^2_{\mathbf r'}
\Phi(\mathbf r')dV'=-4\pi\Phi(\mathbf r),
\end{equation}
Similar reasoning can be performed for each coordinate, to come to the following relation for the vector quantity:

\begin{equation}
\label{1.6}
\int\frac{1}{|\mathbf r-\mathbf r'|}\nabla^2_{\mathbf r'}
\mathbf V(\mathbf r')dV'
=-4\pi\mathbf V(\mathbf r).
\end{equation}

The meaning behind these expressions lies in that the scalar, $\Phi(\mathbf r)$, or vector, $\mathbf V(\mathbf r)$ functions at a point $\mathbf r$ are fully governed by the integral over the whole space of the Laplacian of the same functions taken with the weight $\frac{-1}{4\pi|\mathbf r-\mathbf r'|}$.

\section{Application of the formulas to an electric field}

Eq. (\ref{1.6}) was derived for potential  vectors. This follows already from a straightforward observation that the initial Eq. (\ref{1.1}) was derived from the gradient of a scalar, which accounts for the appearance of potential vectors only (see, for instance, \cite{2}, Ch. 1). Therefore it can be applied to potential fields. Let us apply the above equalities to a constant electric field. We shall use the Gaussian absolute system of units (in this case the expressions derived do not acquire any coefficients). For function $\Phi$ we choose the scalar electric potential $\varphi$. Potential $\varphi$ satisfies the Poisson's equation ($\rho(\mathbf r')$ is the charge distribution):

\begin{equation}
\label{2.1}
\nabla^2_{\mathbf r'}\varphi(\mathbf r')=-4\pi\rho(\mathbf r').
\end{equation}
Substituting expression (\ref{2.1}) into relation (\ref{1.5}) yields the well known expression for the scalar potential:

\begin{equation}
\label{2.2}
\varphi(\mathbf r)=
\int\frac{\rho(\mathbf r')dV'}{|\mathbf r-\mathbf r'|}.
\end{equation}

We have not, however, obtained here anything new. 

Let us apply the equalities derived by us to the  constant electric field. The constant electric field  satisfies Maxwell’s equations

\begin{equation}
\label{2.5}
\nabla_{\mathbf r'}\cdot\mathbf E(\mathbf r')=4\pi\rho(\mathbf r'), \qquad
\nabla_{\mathbf r'}\times\mathbf E(\mathbf r')=0.
\end{equation}
We take the gradient from the first equation in (\ref{2.5}). In view of the second equation in (\ref{2.5}), we come to

\begin{equation}
\label{2.6}
\nabla^2_{\mathbf r'}\mathbf E(\mathbf r')=4\pi\nabla_{\mathbf r'}\rho(\mathbf r').
\end{equation}
Invoking now Eq. (\ref{1.6}) we come to the final result

\begin{equation}
\label{2.7}
\mathbf E(\mathbf r)=
-\int\frac{1}{|\mathbf r-\mathbf r'|}\nabla_{\mathbf r'}
\rho(\mathbf r') dV'.
\end{equation}

\section{Correspondence between the proposed and classical formulas}

We readily see that Eq. (\ref{2.7}) does not coincide with the well-known expression for the electric field

\begin{equation}
\label{3.1}
\mathbf E(\mathbf r)=
\int\frac{\mathbf r-\mathbf r'}{|\mathbf r-\mathbf r'|^3}
\rho({\mathbf r'})dV'.
\end{equation}
In Eqs. (\ref{2.7}) and (\ref{3.1}) neither the integrands nor even the directions of the vectors $(\mathbf r-\mathbf r')$ and 
$\nabla_{\mathbf r'}\rho(\mathbf r')$ coincide. Nevertheless, both expressions describe the same electric field $\mathbf E(\mathbf r)$. This can be demonstrated easily.

\begin{equation}
\label{3.2}
\mathbf E(\mathbf r)=-\nabla_{\mathbf r}\varphi(\mathbf r)=
-\int\nabla_{\mathbf r}\frac{1}{|\mathbf r-\mathbf r'|}
\rho(\mathbf r') dV'.
\end{equation}
The operator $\nabla_{\mathbf r}$ may  be introduced under the integral sign, because the differentiation and the integration are performed here over different coordinates, $\mathbf r$ and $\mathbf r'$. Differentiation will yield Eq. 
(\ref{3.1}). On the other hand, replacing by the Eq. (\ref{1.3}) differentiation over coordinates $\mathbf r$ with that over coordinates $\mathbf r'$ yields

\begin{equation}
\label{3.3}
\mathbf E(\mathbf r)=
\int\left(\nabla_{\mathbf r'}\frac{1}{|\mathbf r-\mathbf r'|}\right)
\rho(\mathbf r')dV'.
\end{equation}
This integral can be taken by parts. Because the integration is performed over the whole space, the surface integral over an infinitely remote surface vanishes, we come to Eq. (\ref{2.7}). This proves conclusively that the expressions described by equalities (\ref{2.7}) and (\ref{3.1}) coincide. Besides, we obtain an equality 

\begin{equation}
\label{3.4}
\int
\rho(\mathbf r')\nabla_{\mathbf r'}\frac{1}{|\mathbf r-\mathbf r'|} dV'=
-\int\frac{1}{|\mathbf r-\mathbf r'|}\nabla_{\mathbf r'}
\rho(\mathbf r') dV'.
\end{equation}

Now what is the physical meaning of Eqs. (\ref{2.2}) and (\ref{2.7})?

The physical meaning of Eq. (\ref{2.2}) appears clear; indeed, every element of the charge $dq(\mathbf r')=\rho(\mathbf r') dV'$ generates a perturbation in space which falls off inversely proportional to the distance from this element. Eq. 
(\ref{2.2}) describes the total perturbation at point $\mathbf r$ due to all the charges. We call this total perturbation the scalar potential $\varphi(\mathbf r)$.
 Eq. (\ref{2.7}) has exactly the same meaning. In this case, however, the perturbation of space should be identified not with the charge but with the charge gradient (taken with the opposite sign). Just as in the case of the charge, this perturbation produced by the charge gradient falls off inversely proportional to the distance from this element. The total perturbation at point $\mathbf r$ from all elements of the charge gradient is the electric field $\mathbf E(\mathbf r)$ (which in itself is a potential gradient (taken with the opposite sign)).

Assume the whole space to be filled uniformly by charges. It is intuitively obvious that the electric field should in this case be zero, notwithstanding the total number of charges being enormous. Eq. (\ref{2.7}) substantiates this conclusion: if there is no charge gradient, electric field must be zero. 

As already mentioned, the appearance of the potential $\varphi(\mathbf r)$ at point $\mathbf r$ may be assigned to combined action of all the charges present in the whole space.

By analogy with hydrodynamics, starting from the equality 
$\nabla\cdot\mathbf E(\mathbf r)=4\pi\rho(\mathbf r)$, the charge is assumed to be the source of the field $\mathbf E(\mathbf r)$. Said otherwise, the charge located at a given point generates a quantity 
$\nabla\cdot\mathbf E(\mathbf r)$ at the same point. Eq. (\ref{2.7}) suggests, however, that the appearance of the field $\mathbf E(\mathbf r)$ should rather be associated with the total effect generated by the {\itshape\bfseries gradient} of all charges distributed over the whole space.

\section{Relation with generalized functions}

Eq. (\ref{2.7}) may be considered as a functional and, hence, one can employ laws obeyed by functionals and generalized functions (see, for instance, 
\cite{4} \$ 6). Consider the general expression for a derivative of generalized functions

\begin{equation}
\label{4.1}
\int (D^{\alpha}f(x))\Psi(x)dx=(-1)^{|\alpha|}\int f(x)D^{\alpha}
\Psi(x)dx,
\end{equation}
or, in the symbolic form (in notations of \cite{4})

\begin{equation}
\label{4.2}
(D^{\alpha}f,\Psi)=(-1)^{|\alpha|}(f,D^{\alpha}\Psi).
\end{equation}
Here $D^{\alpha}$ is the differential operator, and $f(x)$ and $\Psi(x)$ are the principal and generalized functions, respectively. Eq. (\ref{4.1}) indicates that the derivative operator can be switched over from the principal to the generalized function and back. An analysis of Eqs. (\ref{1.2}) and (\ref{3.4}) suggests that they are actually a particular case of the general equality
 (\ref{4.1}) for $\alpha=2$ and $\alpha=1$ (in which case $\Phi(\mathbf r')$
 and $\rho(\mathbf r')$ represent the principal functions, and 
$1/{|\mathbf r-\mathbf r'|}$ the generalized function). Eq. (\ref{4.1}) for scalar quantities can be recast in the following form

\begin{equation}
\label{4.3}
\int (\nabla^{\alpha}_{\mathbf r'}f(\mathbf r'))\Psi(\mathbf r')dV'=
(-1)^{|\alpha|}\int f(\mathbf r')
\nabla^{\alpha}_{\mathbf r'}\Psi(\mathbf r')dV',
\end{equation}
where with $f(\mathbf r')$ one identifies the principal functions 
$\Phi(\mathbf r')$ or $\rho(\mathbf r')$, and with $\Psi(\mathbf r')$, the generalized function $\frac{1}{|\mathbf r-\mathbf r'|}$. The gradient is taken over the $(\mathbf r')$ coordinates. 

We rewrite now Eq. (\ref{1.6}) in the form

\begin{equation}
\label{4.4}
\int\frac{1}{|\mathbf r-\mathbf r'|}\nabla^2_{\mathbf r'}
\mathbf V(\mathbf r')dV'
=\int\mathbf V(\mathbf r')\nabla^2_{\mathbf r'}
\frac{1}{|\mathbf r-\mathbf r'|}dV'.
\end{equation}
Any potential vector can be regrouped in an obvious equality

\begin{equation}
\label{4.5}
\nabla_{\mathbf r'}\cdot\left(\frac{1}{|\mathbf r-\mathbf r'|}\cdot
\mathbf V(\mathbf r')\right)=\frac{1}{|\mathbf r-\mathbf r'|}
\nabla_{\mathbf r'}\cdot\mathbf V(\mathbf r')+
\mathbf V(\mathbf r')\cdot\nabla_{\mathbf r'}
\frac{1}{|\mathbf r-\mathbf r'|},
\end{equation}
Integration now over the whole space, combined with carrying out all the procedures outlined above, leads us to

\begin{equation}
\label{4.6}
\int\frac{1}{|\mathbf r-\mathbf r'|}
\nabla_{\mathbf r'}\cdot\mathbf V(\mathbf r')dV'=
-\int\mathbf V(\mathbf r')\cdot\nabla_{\mathbf r'}
\frac{1}{|\mathbf r-\mathbf r'|}dV'.
\end{equation}
Turning now to Eqs. (\ref{4.4}) and (\ref{4.6}), we see that they may be considered as a particular case of Eq. (\ref{4.1}) for $\alpha=2$ and 
$\alpha=1$:

\begin{equation}
\label{4.7}
\int (\nabla^{\alpha}_{\mathbf r'}\cdot\mathbf V(\mathbf r'))\Psi
(\mathbf r')dV'=
(-1)^{|\alpha|}\int \mathbf V(\mathbf r')\cdot
\nabla^{\alpha}_{\mathbf r'}\Psi(\mathbf r')dV'.
\end{equation}
Here, as before, $\Psi(\mathbf r')=\frac{1}{|\mathbf r-\mathbf r'|}$, and for the function $\mathbf V(\mathbf r')$ one may take any potential function, for instance, the electric field $\mathbf E(\mathbf r')$.

\section{Example of electric field calculation}

We are going to illustrate the use of Eq. (\ref{2.7})  for electric field calculation in the simplest case. We choose for this purpose a spherically symmetric charge distribution

\begin{equation}
\label{5.1}
\rho(\mathbf r)=\rho(r)=Ae^{-\frac{2r}{a}}.
\end{equation}
This distribution coincides with the charge distribution of electron in the ground state of a hydrogen atom. In normalizing this distribution to the electron charge, coefficient $A$ has the value $A=-e_0/\pi a^3$. Here $e_0$ is the absolute magnitude of the electron charge, and $a$ is the Bohr radius. Because of the charge being spherically symmetric, the electric field of this distribution will have only one component $\mathbf E(\mathbf r)=E_r$ along the $\mathbf r$ axis. The electric field of this distribution was calculated by standard classical methods (see, e.g., Problem 83 in \cite{5}):

\begin{equation}
\label{5.2}
\mathbf E(\mathbf r)=\left[-\frac{e_0}{r^2}+\frac{e_0}{r^2}
e^{-\frac{2r}{a}}\left(2\frac{r^2}{a^2}+
2\frac{r}{a}+1\right)\right]\mathbf n_r,
\end{equation}
where $\mathbf n_r$ is the unit vector along the $\mathbf r$ axis.

Calculate now the electric field produced by distribution (\ref{5.1}) with the use of Eq. (\ref{2.7}). The charge gradient is

\begin{equation}
\label{5.3}
\nabla\rho(\mathbf r)=-\frac{2}{a}Ae^{-\frac{2r}{a}}\mathbf n_r=
-\frac{2}{a}\rho(r)\mathbf n_r.
\end{equation}
The charge gradient (\ref{5.3}) has also only one component along the 
$\mathbf r$ axis. To calculate integral (\ref{2.7}),  we can use an expansion in spherical harmonics. This expansion, presented in the form most convenient for us, can be found in \cite{5}, Ch. 2. The charge gradient 
(\ref{5.3}) is, however, a vector quantity, while the distribution is given for a scalar quantity. (Most frequently this expansion is employed in determination of the $\varphi(\mathbf r)$ potential from a known charge distribution 
$\rho(\mathbf r)$).
To calculate the integral of a vector quantity, we invoke the procedure proposed in \cite{6}. This paper describes how one can use the symmetry of a vector distribution to reduce integration of a vector function to that of one scalar function. We are going to employ the same approach.

Because our distribution of charge, charge gradient and, hence, of the electric field has spherical symmetry, it will suffice to determine the electric field only on one arbitrary straight line passing through the charge center. The most convenient approach in the spherical coordinate system $r, \vartheta, \varphi$ is to determine the field on the $\vartheta=0$ axis (as a matter of convenience, we are going to call it the $Z$ axis).
The coordinates of a point on the $Z$ axis are $(r,0)$. The third coordinate 
$\varphi$ on the $\vartheta=0$ axis remains indeterminate, but in our case nothing depends on the $\varphi$ coordinate. But then the field at any point $\mathbf r$ will be $\mathbf E(\mathbf r)=E(r,0)\mathbf n_r$. Expand vector (\ref{5.3}) into two components: along the $Z$ axis and perpendicular to it. In integration, for two vectors at points $r, \vartheta, \varphi$ and $r, \vartheta, \varphi+\pi$, the components along the $Z$ axis will add, while the perpendicular ones will be subtracted. Incidentally, both these points are located at the same distance from a point $(r,0)$ on the $Z$ axis.
Because all perpendicular components will vanish after integration, it appears sufficient to include in integration only the components parallel to the $Z$ axis. This implies that in place of integration of vectors we can integrate the scalar projections of vectors on the $Z$ axis. These projections can be written as

\begin{equation}
\label{5.4}
d(r,\vartheta)=-\frac{2}{a}\rho(r)\cos(\vartheta)=
-\frac{2}{a}Ae^{-\frac{2r}{a}}\cos(\vartheta).
\end{equation}
This expression can be written in a different way

\begin{equation}
\label{5.5}
d(r,\vartheta)=-\frac{2}{a}Ae^{-\frac{2r}{a}}\cos(\vartheta)=
-\frac{2}{a}Ae^{-\frac{2r}{a}}\sqrt{\frac{4\pi}{3}}Y_{10},
\end{equation}
where $Y_{10}=\sqrt{\frac{3}{4\pi}}\cos(\vartheta)$ is a spherical function of order $1,0$ (e.g. \cite{5} in Appendix 2, or \cite{7}, \S~1). To determine the field at axis $Z$, i.e. the quantity $E(r,0)$, it is this expression (\ref{5.5}) that should be inserted into the integrals to find the multipole moments. The expressions proposed for calculation of the field $E(r,0)$ and the multipole moments will look in the following way

\begin{equation}
\label{5.6}
E_Q(r,0)=-\sum_{l=0}^\infty\sum_{m=-l}^l\sqrt{\frac{4\pi}{2l+1}}\cdot
\frac{Q_{lm}(r)Y_{lm}(\vartheta, \varphi)}{r^{l+1}}  \qquad (r>r'),
\end{equation}
where $Q_{lm}(r)$ is a multipole moment of order $l, m$:

\begin{equation}
\label{5.7}
Q_{lm}(r)=\sqrt{\frac{4\pi}{2l+1}}\int_0^rd(r',\vartheta')r'^lY_{lm}^*
(\vartheta',\varphi')dV',
\end{equation}

\begin{equation}
\label{5.8}
E_G(r,0)=-\sum_{l=0}^\infty\sum_{m=-l}^l\sqrt{\frac{4\pi}{2l+1}}
r^l{G_{lm}(r)Y_{lm}(\vartheta, \varphi)}  \qquad (r<r'),
\end{equation}
where 

\begin{equation}
\label{5.9}
G_{lm}(r)=\sqrt{\frac{4\pi}{2l+1}}\int_r^\infty\frac{d(r',\vartheta')}
{r'^{l+1}}Y_{lm}^*(\vartheta',\varphi')dV'.
\end{equation}
The minus signs $(-)$ before the double sums in Eqs. (\ref{5.6}) and 
(\ref{5.8}) are related to the $(-)$ sign in Eq. (\ref{2.7}).

Functions $Q_{lm}(r)$ and $G_{lm}(r)$ depend on $r$ in the upper and lower limits of integration as on a parameter.

The field at a given point $(r,0)$ on the $Z$ axis is a sum of the fields calculated for $r>r'$ and $r<r'$:  $E(r,0)=E_Q(r,0)+E_G(r,0)$.

The orthogonality of the $Y_{lm}$  spherical functions leaves after calculation of the multipole moments only two of them, $Q_{10}$ and $G_{10}$, and of the whole series only two terms will be left, one term for $r>r'$, and another for $r<r'$. (If functions (\ref{5.1}) and (\ref{5.3}) had axial rather than spherical symmetry, we would have obtained a single series containing the multipole moments $Q_{l0}$ and $G_{l0}$). The calculations yield the following values  for 
the multipole  moments

\begin{equation}
\label{5.10}
Q_{10}(r)=e_0\cdot\frac{4}{3}\left[e^{-\frac{2r}{a}}
\left(-\frac{r^3}{a^3}-\frac{3}{2}\cdot\frac{r^2}{a^2}-
\frac{3}{2}\cdot\frac{r}{a}-\frac{3}{4}\right)+
\frac{3}{4}\right],
\end{equation}

\begin{equation}
\label{5.11}
G_{10}(r)=\frac{e_0}{a^3}\cdot\frac{4}{3}e^{-\frac{2r}{a}}.
\end{equation}
All the other multipole moments are zero.

In substitution of the miltipole moments (\ref{5.10}) and (\ref{5.11}) into the equations for the field, (\ref{5.6}) and (\ref{5.8}), it should be kept in mind that what we are looking for is only the field on $Z$ axis, i.e., on the 
$\vartheta=0$ axis. Therefore the quantity $\vartheta$ in Eqs. (\ref{5.6}) and (\ref{5.8}) should be given the value zero. This all yields $cos{\vartheta}=1$, and the field on the $Z$ axis, $E(r,0)=E_Q(r,0)+E_G(r,0)$, will acquire the form 

\begin{equation}
\label{5.12}
E(r,0)=-\frac{e_0}{r^2}+\frac{e_0}{r^2}
e^{-\frac{2r}{a}}\left(2\frac{r^2}{a^2}+
2\frac{r}{a}+1\right).
\end{equation}
The electric field at any point is 
$\mathbf E(\mathbf r)=E(r,0)\mathbf n_r$, which coincides with expression 
(\ref{5.2}). To sum up, using Eq. (\ref{2.7}), we have come to the same expression that is obtained with classical methods.

\end{document}